\documentclass[12pt]{article} 
\usepackage{amsmath, amsfonts, amssymb}
\usepackage[margin=0.75in]{geometry}
\usepackage{graphicx}
\usepackage{hyperref}
\usepackage{setspace}
\usepackage{color}
\usepackage{algorithm} 
\usepackage{algpseudocode}
\usepackage{tabularx}
\usepackage{lscape} %vertical table
\usepackage{rotating}
\usepackage{hyphenat}
\usepackage{enumitem}
\usepackage{bm}
\usepackage{placeins} %Added by Ali, to restrict images inside sections/subsections by ``\FloatBarrier'' code
\usepackage{color} %Added by Ali, to use another color in the text! with code \textcolor{color_name}
\usepackage{amsmath}
\usepackage{setspace}
\usepackage[capitalise]{cleveref}
\usepackage{fancyhdr}
\usepackage{diagbox}

\usepackage{soul}
\usepackage{cite} %group citations together

\newcommand{\edit}[1]{\color{black}#1 \color{black}}

\title{Integrating multi-fidelity blood flow data with reduced-order data assimilation }
\author{Milad Habibi$^1$\and Roshan M. D'Souza$^2$ \and Scott T. M. Dawson$^3$\and Amirhossein Arzani$^1$}

\date{}

\begin{document}

\maketitle

%\newsavebox{\astrutbox}
%\sbox{\astrutbox}{\rule[-5pt]{0pt}{20pt}}
%\newcommand{\astrut}{\usebox{\astrutbox}}

\begin{center}
$^1$Department of Mechanical Engineering, Northern Arizona University, Flagstaff, AZ, United States \\
$^2$Department of Mechanical Engineering, University of Wisconsin--Milwaukee, Milwaukee, WI, United States.\\
$^3$Department of Mechanical, Materials, and Aerospace Engineering, Illinois Institute of Technology, Chicago, IL, United States\\
\end{center}

\bigskip

\noindent Correspondence:\\
Amirhossein Arzani,\\
Northern Arizona University,\\
Flagstaff, AZ, 86011\\
Email: amir.arzani@nau.edu
%\newpage

\thispagestyle{empty}

%\doublespacing
\begin{abstract}
High-fidelity patient-specific modeling of cardiovascular flows and hemodynamics is challenging. Direct blood flow measurement inside the body with in-vivo measurement modalities such as 4D flow magnetic resonance imaging (4D flow MRI) suffer from low resolution and acquisition noise. In-vitro experimental modeling and patient-specific computational fluid dynamics (CFD) models are subject to uncertainty in patient-specific boundary conditions and model parameters. Furthermore, collecting blood flow data in the near-wall region (e.g., wall shear stress) with experimental measurement modalities poses additional challenges. In this study, a computationally efficient data assimilation method called reduced-order modeling Kalman filter (ROM-KF) was proposed, which combined a sequential Kalman filter with reduced-order modeling using a linear model provided by dynamic mode decomposition (DMD). The goal of ROM-KF was to overcome low resolution and noise in experimental and uncertainty in CFD modeling of cardiovascular flows. The accuracy of the method was assessed with 1D Womersley flow, 2D idealized aneurysm, and 3D patient-specific cerebral aneurysm models. Synthetic experimental data were used to enable direct quantification of errors using benchmark datasets. The accuracy of ROM-KF in reconstructing near-wall hemodynamics was assessed by applying the method to problems where near-wall blood flow data were missing in the experimental dataset. The ROM-KF method provided blood flow data that were more accurate than the computational and synthetic experimental datasets and improved near-wall hemodynamics quantification.


\end{abstract}
\newpage

\section{Introduction} \label{sec:intro}

One of the major challenges in transforming blood flow dynamics (hemodynamics) modeling into clinical practice is overcoming modeling limitations. Despite our improved understanding of disturbed blood flow patterns in diseased arteries~\cite{ArzaniShadden12,ArzaniShadden18,Asgharzadehetal19,Kheradvaretal19} and accumulating evidence for the role of hemodynamics in regulating cardiovascular disease~\cite{Pedrigietal16,Timminsetal17}, we do not yet possess a high-fidelity technology (with limited uncertainty and error) for the quantification of spatiotemporally varying hemodynamic parameters such as wall shear stress (WSS). Leading modalities for modeling 3D time-resolved hemodynamics include patient-specific computational fluid dynamic (CFD), 4-dimensional magnetic resonance imaging (4D flow MRI), and particle image/tracking velocimetry (PIV/PTV). However, all of these modalities suffer from different limitations. Recent ultra-high-resolution CFD simulations have leveraged high-performance computing and minimally dissipative CFD solvers for high spatiotemporal resolution modeling of hemodynamics with minimal numerical errors~\cite{KhanValenSteinman15,Arzani18}. Advances in experimental measurement techniques \edit{such as} PIV and PTV have reduced experimental measurement errors, albeit measurement noise and difficulty in measuring near-wall flow remains a challenge~\cite{Raffeletal18,Brindiseetal19}. Perhaps the most \edit{prominent} issue with these modeling techniques is the uncertainty in patient-specific modeling parameters. Inlet and outlet boundary conditions are not always precisely known and blood rheology variability and uncertainty in patient-specific tissue material properties (in a deformable wall setting) limit the utility of even ultra-high-resolution models once they are interpreted for patient-specific applications. 

Direct in-vivo measurement of hemodynamics using 4D flow MRI does not require boundary conditions or assumptions \edit{about} constitutive models; however, it suffers from several inherent limitations such as low spatiotemporal resolution and measurement noise~\cite{Barkeretal12,Cibisetal16,Roloffetal19}.  All of these limitations and constraints \edit{raise} this question: can we obtain more accurate hemodynamics data from low resolution noisy experimental data and high-resolution computational data with uncertain parameters? Another limitation of in-vitro and in-vivo experimental techniques is in the quantification of near-wall flow. Near-wall hemodynamic measures such as WSS are difficult to quantify accurately using experimental techniques. Is it possible to improve near-wall hemodynamics data by leveraging measurements away from the wall?

Data-driven modeling techniques such as data assimilation (DA) have the potential to overcome the limitations and challenges in obtaining high fidelity hemodynamics data~\cite{ArzaniDawson20}. DA refers to a group of uncertainty quantification and data fusion techniques that merge information from numerical/mathematical models and measurement data to determine a more accurate state of the system~\cite{Aschetal16}. The mathematical model is used to advance the solution forward in time, while the measurement data improves the model prediction and provides the means to mitigate uncertainty. DA solves the problem of data fusion by leveraging prior knowledge about the error distribution in the mathematical model and measurement data \edit{and} appropriately combines the two for achieving superior accuracy. DA could be used to estimate unknown model parameters (e.g., boundary or initial conditions) or improve data fidelity by combining multi-modality data (e.g., computational and experimental). 

%Kalman filter is employed in many applications such as weather forecasting~\cite{Navon09}, navigation~\cite{Hasanetal09}, economics\cite{Augeretal13}, et

DA is divided into variational and sequential methods~\cite{Aschetal16}. In the variational approach, an adjoint model needs to be formulated and solved, which is a computationally expensive approach. Kalman filter as a sequential DA approach is a recursive algorithm that estimates the state of a system by using the model of uncertainties, measurement noise, and process noise~\cite{Aschetal16}. In different applications, the Kalman filter is \edit{extended} to multiple types such as the unscented Kalman filter~\cite{WanVan00}, the extended Kalman filter~\cite{HoshiyaSaito84}, and the ensemble Kalman filter~\cite{Evensen03} (among others). In the cardiovascular fluid mechanics community, DA has been used in different problems. A popular application is using DA to estimate the boundary conditions needed in CFD simulations (e.g., lumped parameter networks) ~\cite{Pantetal14,Canutoetal20,Arthursetal20}. Merging 4D flow MRI and CFD data using variational~\cite{Funkeetal19} and sequential~\cite{Gaidziketal19,Gaidziketal20} DA approaches is another trending research area. Marching large nonlinear hemodynamic systems forward in time in DA modeling is a major challenge. Unscented Kalman filter is proposed to overcome the nonlinearity issue~\cite{WanVan00}; however, it requires several sampling of large nonlinear systems. The ensemble Kalman filter method is proposed to solve the computational storage problem of large systems; however, it relies on random Monte Carlo sampling and therefore has \edit{a} high computational cost~\cite{Gaidziketal19}. 

A major issue in most DA approaches is the high computational cost. An appropriate combination of reduced-order modeling (ROM) and DA is a promising approach for merging multi-modality data.  Reduced-order unscented Kalman filter is used in estimating different parameters needed in multi-physics cardiovascular biomechanics simulations~\cite{Bertoglioetal12,Corradoetal15,Arthursetal20}. However, these studies have mainly performed parameter estimation. Multi-fidelity data fusion using reduced-order Kalman filter has received less attention. Recent advances in deep learning have motivated the development of autoencoders \edit{for} reduced-order data assimilation with neural networks~\cite{Amendolaetal20,Macketal20,Arcuccietal21}; however, their success in complex cardiovascular flows remains to be investigated. Another approach is to develop a data-driven ROM to enable a computationally efficient framework for marching the solution forward in time.  Specifically, Galerkin proper orthogonal decomposition (POD) has been used to develop a reduced-order Kalman filter model and address the computational cost issue (see~\cite{Hayase15} and references therein). However, Galerkin POD is an intrusive reduced-order approach and is subject to stability issues in advection dominated flows~\cite{Romainetal14,Ahmedetal20}.

Dynamic mode decomposition (DMD) is a more recent ROM approach~\cite{Schmid10,Tairaetal17}. DMD is non-intrusive (equation-free) and provides a simple linear dynamical system model (${\mathbf{x}_{k+1}}=\mathbf{A}\mathbf{x}_{k}$), which could be easily integrated in forward time. DMD has been used extensively in different applications, and its mathematical simplicity has facilitated several extensions~\cite{Williamsetal15,Proctoretal16,Dawsonetal16}. In the context of cardiovascular flows, DMD has been applied to blood flow problems~\cite{DiKadem19,Le21} and multi-stage DMD has been proposed for studying blood flow physics during different phases of the cardiac cycle~\cite{Habibietal20}. The Kalman filter could be easily applied to any linear dynamical system. DMD provides this linear system in a reduced-order space and therefore seems to be a promising approach for developing reduced-order DA models. There are extensions of DMD that incorporate Kalman filter based ideas for parameter estimation~\cite{Nonomuraetal18} and denoising~\cite{Nonomuraetal19}. Recently, DMD has been used in combination with Kalman filtering and smoothing for denoising time-resolved hemodynamics data~\cite{Fathietal20}. Additionally, DMD is used for computing the mapping matrix between low-resolution and superresolved 4D flow MRI data~\cite{Perezetal20}. However, none of these studies have performed a DA study where low-resolution experimental and uncertain CFD data are combined to address not only low resolution and noise in experimental data but also uncertainty in computational modeling. 

%However, none of these studies can be considered a DA study for addressing challenges to obtain high-fidelity data

In this study, we propose a reduced-order model Kalman filter (ROM-KF) \edit{approach} leveraging DMD as an efficient data-driven linear operator. Our goal is to combine time-resolved multi-fidelity hemodynamics data using ROM-KF to produce more accurate results. Additionally, we would like to see if we can improve near-wall hemodynamics data by only leveraging experimental data away from the wall. This problem is motivated by two observations: 1- In cardiovascular fluid mechanics, we are interested in near-wall hemodynamics and WSS parameters due to their \edit{important} role in regulating vascular health~\cite{Arzanietal17,ArzaniShadden18}. 2-  In-vivo and in-vitro experimental measurements have limitations in quantifying near-wall hemodynamics. It is often easier to measure blood flow away from the vessel wall where wall interference is not present. \edit{We demonstrate how our framework could be utilized to solve this problem.} We test our model in 1D Womersley flow, a 2D idealized cerebral aneurysm, and a 3D patient-specific cerebral aneurysm model. We use synthetic experimental data \edit{to} assess the accuracy of our model with respect to a benchmark dataset and discuss the challenges in extending our framework to real-world experimental datasets.

\section{Methods} \label{sec:method}
We consider three test case models in evaluating our proposed method. 1D blood flow based on Womersley's analytical solution, 2D pulsatile blood flow in an idealized aneurysm, and 3D pulsatile blood flow in a patient-specific cerebral aneurysm is considered. The analytical and numerical solutions based on known parameters are considered as ground truth \edit{data} for verification. Uncertain models are considered where \edit{some} parameters are treated as unknown. Finally, synthetic experimental data are \edit{generated} by downsampling the ground truth data and adding noise. The three test cases and the proposed ROM-KF \edit{approach} are explained below.

\subsection{Test case 1: Womersley's analytical solution}\label{sec:WOMER}  
Womersley's solution represents a simple idealized model for pulsatile blood flow. It is an analytical solution for pulsatile, incompressible flow in rigid straight tubes~\cite{Womersley55}. The solution \edit{describes} velocity variation in the radial direction and time based on given Fourier modes of a pressure gradient waveform.
In this example, we create two sets of synthetic experimental datasets with different noise levels to demonstrate how the Kalman filter can produce superior results by prior knowledge of the noise distribution.

\subsubsection{Ground truth data}
\begin{figure}[h!]
\centering
\includegraphics[width=0.9\textwidth,height=\textheight,keepaspectratio]{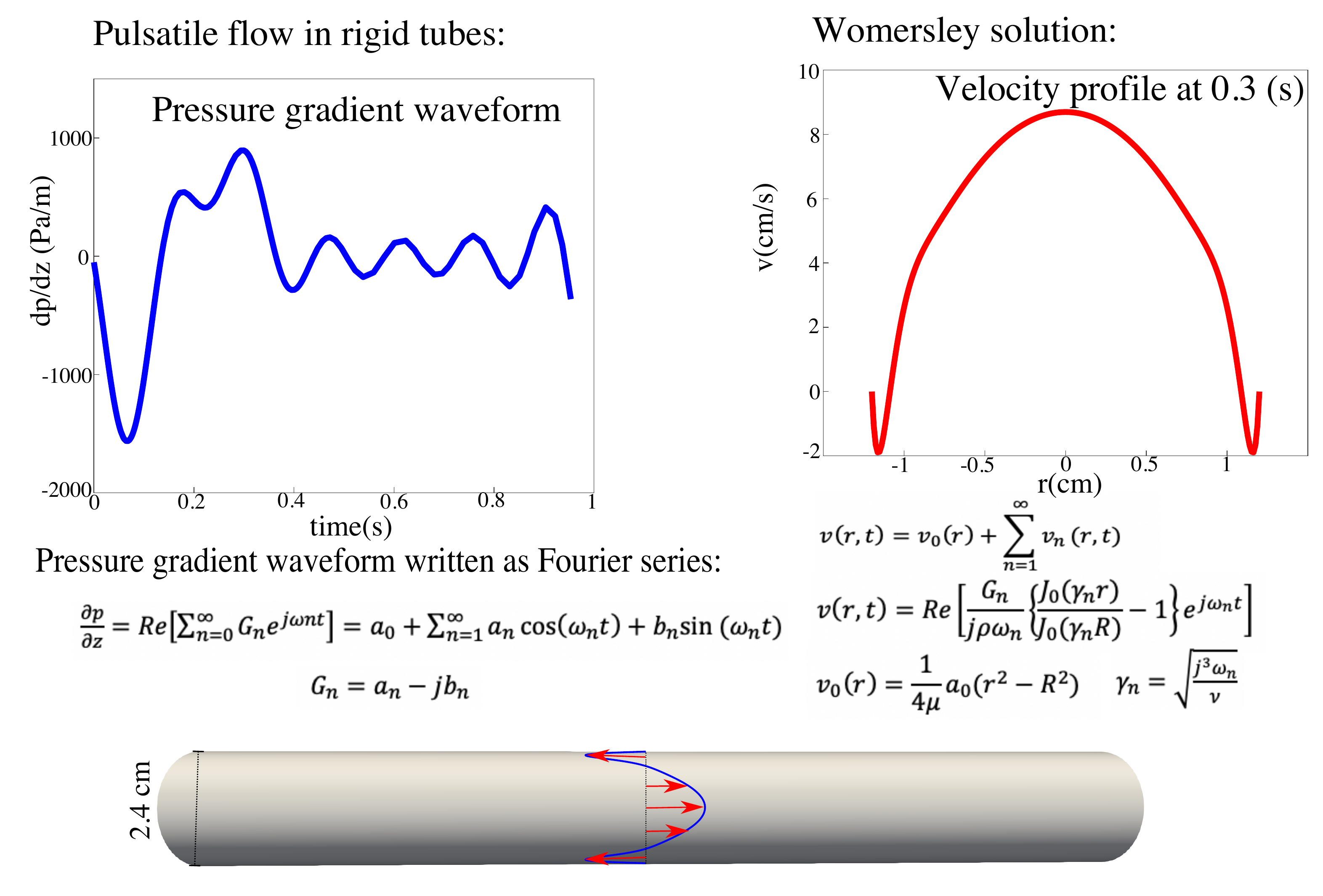}
\caption{The pressure gradient driving the flow  and the corresponding Womersley's solution are shown. The velocity profile is shown at t = 0.3s. The Womersley number is 17.05. The blood density ($\rho$) is 1060 $\frac{kg}{m^3}$ and blood viscosity ($\mu$) is 0.0035 $Pa.s$. }
\label{fig:WOR}
\end{figure}

The pressure gradient waveform and geometric dimensions are shown in Fig~\ref{fig:WOR}, which are taken from a previous study~\cite{Xiaoetal14}. The ground truth velocity dataset is calculated based on the given pressure gradient waveform and using Womersley's analytical solution. This dataset contains noise-free snapshots of velocity. One example snapshot of the data is shown in Fig.~\ref{fig:WOR}.

\subsubsection{Synthetic experimental data}
The first synthetic experimental dataset was created by adding a normally distributed noise with zero mean and 10\% of the maximum velocity as standard deviation. In the second dataset, the standard deviation of noise was set to 20\% of  the maximum velocity. These datasets are referred to as ``lower noise'' and ``higher noise'', respectively.

\subsection{Test case 2: Idealized 2D cerebral aneurysm model}\label{sec:2DCer}  
In this case, an idealized 2D cerebral aneurysm model is considered. The geometry and dimensions of this model are shown in Fig.~\ref{fig:ANE2D}. Here, the problem where there is uncertainty in computational model parameters is considered. Also, the synthetic experimental dataset has low spatial resolution with noise. This test case includes a ground truth dataset with a known inlet boundary condition and viscosity, a computational model with a different inlet boundary condition and viscosity, and a dataset of synthetic noisy experimental data with low spatial resolution. Ground truth, computational, and synthetic experimental datasets are described in the following sections.

\subsubsection{Ground truth data}
Unsteady CFD simulation was performed using the open-source finite element solver FEniCS~\cite{LoggMardalWells12}. Incompressible Navier-Stokes equation were solved with $\rho=1.06 \frac{g}{cm^3}$ and $\mu=0.04$P. The mesh consisted of 15.2K triangular quadratic elements  (equivalent to approximately 60.8K linear elements). \edit{Mesh independence was confirmed by comparing the velocity field to the results obtained from a mesh with 30.3K quadratic elements.}

\begin{figure}[h!]
\centering
\includegraphics[width=\textwidth,height=\textheight,keepaspectratio]{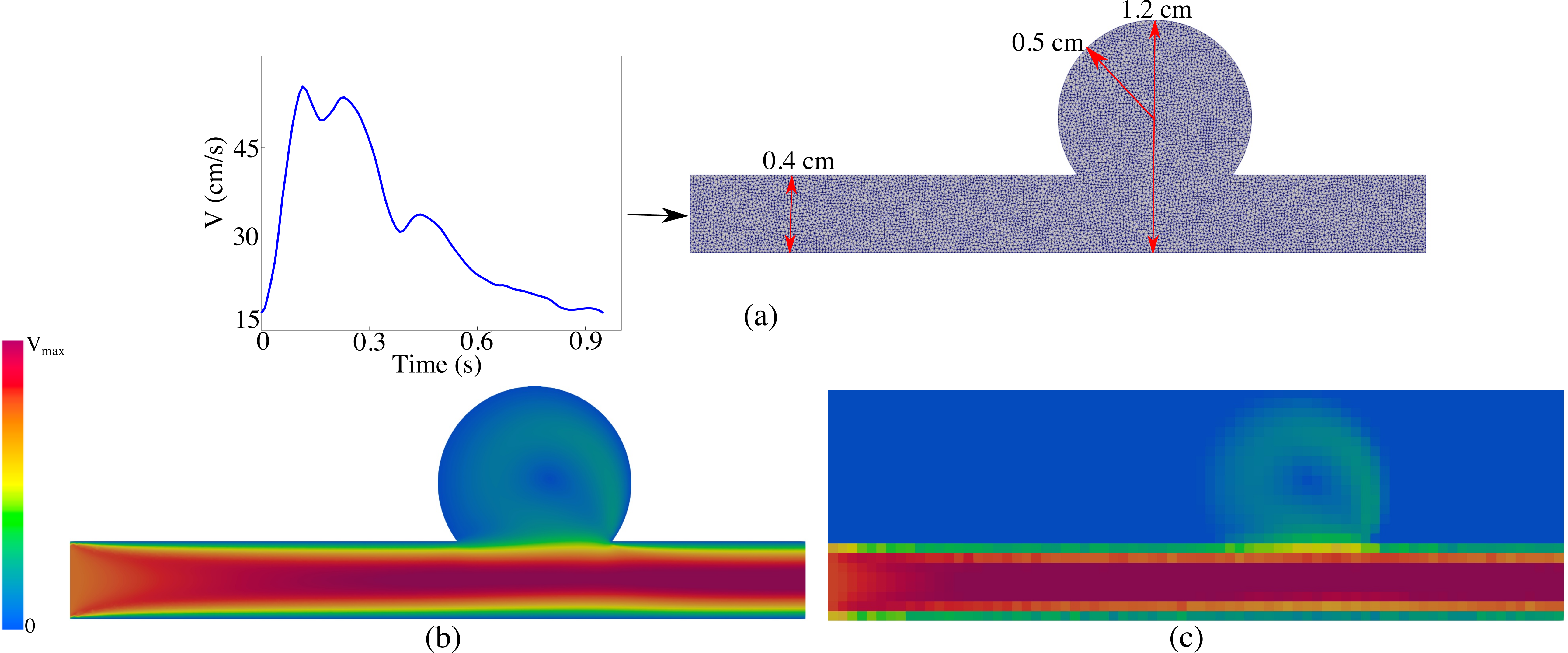}
\caption{(a) The pulsatile waveform used as inlet boundary condition, the geometric dimensions, and the computational mesh are shown. (b) A single snapshot of the velocity profile for the computational data is shown. (c) A single snapshot of the velocity profile for the synthetic experimental data before adding noise is shown. }
\label{fig:ANE2D}
\end{figure}

The velocity profile at the inlet was spatially uniform. \edit{Zero traction was prescribed at the outlet.} The inlet boundary condition waveform is shown in Fig.~\ref{fig:ANE2D}a.

\subsubsection{Computational data}
In the computational data, we assume that we do not have precise knowledge of the inlet boundary condition and viscosity. The inlet boundary condition used is scaled to the 90\% of the ground truth inlet flow waveform and the viscosity used is  $\mu=0.035$P (88\% of the ground truth viscosity). The CFD simulation was performed similarly to the ground truth data. One snapshot of this dataset is shown in Fig.~\ref{fig:ANE2D}b.

\subsubsection{Synthetic experimental data}

4D flow MRI has been used for in-vivo measurement of 3D and time-resolved blood flow velocity data~\cite{Soulatetal20}. However, it suffers from inherent limitations such as measurement noise and low resolution. In this 2D example, we created synthetic experimental data, mimicking some of the 4D flow MRI data features. It should be noted that the synthetic experimental data does not replicate all of the 4D flow MRI features and is highly idealized. Specifically, a rectangle was considered encompassing the 2D aneurysm model and was discretized into square cells ($0.5\times0.5$mm) and the velocity in each cell was calculated based on the averaging over the ground truth velocity nodes located in that cell. Subsequently, a normally distributed noise with zero mean and a standard deviation equal to 10\% of the maximum ground truth velocity was added to create the synthetic data (same standard deviation for all points). One snapshot of this dataset before adding noise is shown in Fig.~\ref{fig:ANE2D}c.

\subsection{Test case 3:  Patient-specific cerebral aneurysm model} \label{sec:3DCer}

An image-based cerebral aneurysm (internal carotid artery aneurysm) model from the online Aneurisk database was selected (Aneurisk ID: C0013) and used in generating the ground truth, computational, and synthetic experimental data as explained below.

\subsubsection{Ground truth data}

\begin{figure}[h!]
\centering
\includegraphics[scale=0.25,keepaspectratio]{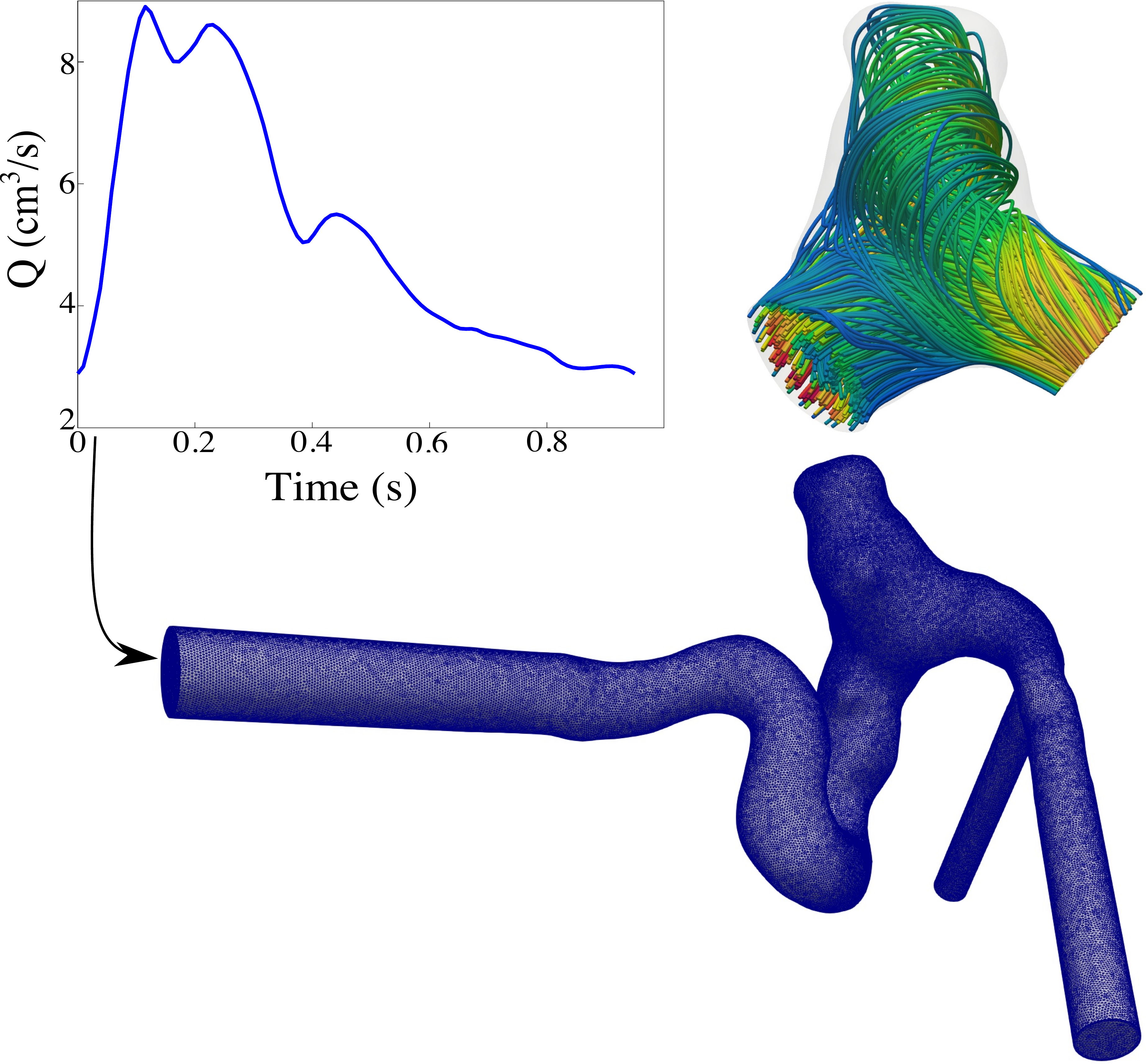}
\caption{The 3D cerebral aneurysm model is shown. The pulsatile waveform used as inlet boundary condition, the 3D mesh, and velocity streamlines for one snapshot are presented.  }
\label{fig:ANE3D}
\end{figure}

The simulation was done using Oasis, an open-source, finite element solver developed in FEniCS~\cite{MortensenValensendstad15}. Rigid walls were assumed, and blood flow was assumed Newtonian~\cite{Arzani18}. To obtain higher accuracy, quadratic tetrahedral elements for velocity and linear elements for pressure  were used~\cite{KhanValenSteinman15}. \edit{The fractional steps method using the incremental pressure correction scheme was used with semi-implicit Adams-Bashforth and Crank-Nicolson for convection and viscous term discretization, respectively~\cite{MortensenValensendstad15}. Oasis has been shown to be ideal for high-resolution simulation of cardiovascular flows~\cite{ValenSteinman14}.} Pulsatile parabolic profile was prescribed at the inlet, and resistance boundary conditions divided the flow at the outlets according to Murray's law. \edit{To reduce the effect of the boundary conditions on the hemodynamics in the aneurysm region, the inlet and outlets were sufficiently extended.} A 6.68M tetrahedral element mesh with three layers of the boundary layer was created in SimVascular. Quadratic elements were used in Oasis for the velocity (equivalent to approximately 53.4M linear elements). \edit{Mesh independence was confirmed by comparing the velocity field to the results obtained from a mesh with 12.07M quadratic elements.} The pulsatile waveform prescribed at the inlet was taken from a previous study~\cite{Hoietal10} and was scaled using scaling laws reported in~\cite{Valenetal15}. The inlet boundary condition is shown in Fig.~\ref{fig:ANE3D}. The viscosity was assumed to be $\mu=0.04$P. Three cardiac cycles were simulated, and the last cardiac was used for data processing. \edit{To obtain high-resolution data,} each cycle was divided into 12,000 time steps.

\subsubsection{Computational data}

In the computational model, we assume uncertainty in the inlet boundary condition and viscosity. The inlet boundary condition is selected to be 90\% of the ground truth data. Also, the viscosity is assumed  $\mu=0.035 $P. 
%One snapshot of this dataset is shown in the Fig.~\ref{fig:ANE3D}. 

\subsubsection{Synthetic experimental data}
The synthetic  experimental dataset is obtained with random spatial downsampling of the ground truth dataset (sampling one point out of every three points) and by adding a normally distributed noise with zero mean and 10\% of the maximum ground truth velocity as the standard deviation (same standard deviation for all points). 

%(synthetic experimental data =Ground truth$+\mathcal{N}(\mathcal{M}=0,\,\sigma=\alpha \%*max(vel))$. 

\subsection{Dynamic mode decomposition (DMD)}

The Kalman filter algorithm requires a discrete linear model for the integration of computational and experimental data. Herein, we propose to use DMD to obtain a linear model for the Kalman filter algorithm. DMD is an equation-free reduced-order model that can find the best linear representation from any given spatiotemporal dataset. Applying DMD on sequential (temporal) snapshots of computational data (N time steps) can provide a discrete linear equation that describes the evolution of the system between any two arbitrary snapshots $k$ and $k+1$ as:

\begin{equation}
\mathbf{x}_{k+1} = \mathbf{A}\mathbf{x}_k\;,
\label{equ:DMD}
\end{equation}
where $\mathbf{A} \in \mathbb{R}^{M\times M}$ is the linear DMD operator and $\mathbf{x}_{k}  \in \mathbb{R}^{M}$  is a snapshot of  the computational data at time-step $k$. Further details about DMD and its variants could be found in~\cite{Kutzetalb16,Tuetal13}. In this work, we used the forward-backward DMD (fbDMD) algorithm, which is more robust to noisy data~\cite{Dawsonetal16}. The DMD algorithm used in this study is explained in the Appendix. Next, we explain how the linear dynamical system model obtained from DMD is used in constructing our Kalman filter and data assimilation algorithm.

\subsection{Kalman filter}

Kalman filter provides the means to improve the estimation of a linear dynamical system with uncertainty/error by leveraging noisy and low-resolution measurement data~\cite{Aschetal16}. The following linear dynamical systems equation describes our computational model as:
\begin{equation}
 \mathbf{x}_{k}=\mathbf{A}\mathbf{x}_{k-1}+\mathbf{w}_{k-1} \;,
 \label{eqn:dynsys}
\end{equation} 
where $\mathbf{x}\in \mathbb{R}^M$ is the state vector, $\mathbf{A}$ is the state transition matrix obtained from our fbDMD algorithm, and $\mathbf{w} \in \mathbb{R}^M$ is a normally distributed white noise process with covariance $\mathbf{Q}$, i.e.\, $\mathbf{w}$$\sim$$\mathcal{N}(0,\mathbf{Q})$. The measurement can be described as:

\begin{equation}
 \mathbf{y}_{k}=\mathbf{H}\mathbf{x}_{k}+\mathbf{v}_{k} \;,
  \label{eqn:exp}
\end{equation} 
where $\mathbf{y}\in \mathbb{R}^P$ is the measurement vector, $\mathbf{x}\in \mathbb{R}^M$ is the state vector given by Eq.~\ref{eqn:dynsys},  $\mathbf{v}\in \mathbb{R}^P$ is the normally-distributed measurement noise with covariance $\mathbf{R}$, i.e., $\mathbf{v}\sim \mathcal{N}(0, \mathbf{R})$, and $\mathbf{H}\in \mathbb{R}^{P\times M}$ is the measurement matrix also known as the observation matrix. The Kalman filter algorithm estimates the state of the system $\mathbf{x}$ given the noisy measurements $\mathbf{y}$ and the noise covariance matrices $\mathbf{Q}$ and $\mathbf{R}$. The Kalman filter algorithm consists of two stages: prediction and correction.  The prediction step is described as follows:

\begin{equation}
%\begin{aligned}[b]
\begin{cases}
{\mathbf{x}_{k}^{(p)}}={\mathbf{A}}{\mathbf{x}}_{k-1}^{(c)} \;, \\
{\mathbf{P}_{k}^{(p)}}={\mathbf{A}}{\mathbf{P}_{k-1}}^{(c)}{\mathbf{A}^{T}}+{\mathbf{Q}} \;.
%\end{aligned}
\end{cases}
\end{equation}

Subsequently, the correction step is applied:
\begin{equation}
%\begin{aligned}[b]  
\begin{cases}
{\mathbf{K}_{k}}={\mathbf{P}}_{k}^{(p)}{\mathbf{H}}^{T}(\mathbf{H}{\mathbf{P}_{k}}^{(p)}\mathbf{H}^{T} +\mathbf{R})^{-1} \;,\\
{\mathbf{x}_{k}}^{(c)}={\mathbf{x}_{k}}^{(p)}+{\mathbf{K}_{k}}(\mathbf{y}_{k}-\mathbf{H}\mathbf{x}_{k}^{(p)})  \;, \\
{\mathbf{P}_{k}^{(c)}}=(\mathbf{I}-\mathbf{K}_{k}\mathbf{H})\mathbf{P}_{k}^{(p)} \;,
%\end{aligned}
\end{cases}
\end{equation}
where $\mathbf{P}$ represents the estimation error covariance and $\mathbf{K}$ denotes the Kalman gain. The superscripts $p$ and $c$ denote predicted and corrected (updated) estimates, respectively.

\subsection{Reduced-order model Kalman filter (ROM-KF)} 

%\edit{Projection of the high-dimensional data onto space spanned by basis vectors (columns of $\mathbf{\tilde{U}}_{r}$) which are also called POD basis vectors, will reduce the dimensional of the data to the number of the basis vectors}.

Representing nonlinear systems such as blood flow in arteries with a linear system might not be trivial. Additionally, the application of the above Kalman algorithm on large systems is computationally expensive due to the matrix inversions involved. Herein, we propose to use fbDMD to provide a linear representation of our system and map our large-scale computational hemodynamics data onto a reduced-order basis approximating the  computational data basis ($\mathbf{\tilde{U}}_{r}$) as computed from our fbDMD algorithm explained in the Appendix. $\mathbf{\tilde{U}}_{r}$ represents a matrix of proper orthogonal decomposition (POD) modes. It will provide a reduced-order linear dynamical system model, which could be readily incorporated within the above Kalman filter framework. First, a low-dimensional state vector $\mathbf{z}\in \mathbb{R}^r$ is estimated from the computational data using $ \mathbf{x}=\mathbf{\tilde{U}}_{r}\mathbf{z}$ and combining with Eq.~\ref{eqn:dynsys} to get

\begin{equation}
 \mathbf{\tilde{U}}_{r}^{T}\mathbf{x}_{k}=\mathbf{z}_{k}=\mathbf{\tilde{U}}_{r}^{T}\mathbf{A}\mathbf{\tilde{U}}_{r}\mathbf{z}_{k-1}+\mathbf{\tilde{U}}_{r}^{T}\mathbf{w}_{k-1} \;,
\end{equation} 
where the new reduced-order forward model matrix is $\mathbf{A}_{z}=\mathbf{\tilde{U}}_{r}^{T}\mathbf{A}\mathbf{\tilde{U}}_{r}\in \mathbb{R}^{r\times r}$. Similarly, we rewrite the measurement equation (Eq.~\ref{eqn:exp}) using the reduced-order data. Consequently, the computational and measurement models in a reduced-order space could be described as:

 \begin{equation}
 \mathbf{z}_{k}=\mathbf{A}_{z}\mathbf{z}_{k-1}+\mathbf{\tilde{U}}_{r}^{T}\mathbf{w}_{k-1} \;,
 \end{equation} 
 \begin{equation}
 \mathbf{y}_{k}=\mathbf{H}_{z}\mathbf{z}_{k}+\mathbf{v}_{k} \;,
\end{equation} 
where ${\mathbf{H}_{z}}={\mathbf{H}\mathbf{\tilde{U}}_{r}}\in \mathbb{R}^{P\times r}$ is the reduced-order measurement matrix, $\mathbf{\tilde{U}}_{r}^{T}\mathbf{w}_{k-1}$$\sim$$\mathcal{N}(0,\mathbf{Q}_{z})$ is the projected noise, and  $\mathbf{Q}_{z}=\mathbf{\tilde{U}}_{r}^{T}\mathbf{Q}\mathbf{\tilde{U}}_{r}\in \mathbb{R}^{r\times r}$ is the reduced-order noise covariance.

Considering the reduced-order model and measurement equations, the prediction part of the Kalman filter becomes
\begin{equation}
%\begin{aligned}[b]
\begin{cases}
{\mathbf{z}_{k}}^{(p)}={\mathbf{A}_{z}}{\mathbf{z}}_{k-1}^{(c)} \;,  \\
{\mathbf{P}_{k}}^{(p)}={\mathbf{A}_{z}}{\mathbf{P}_{k-1}}^{(c)}{\mathbf{A}_{z}^{T}}+{\mathbf{Q}_{z}} \;.
%\end{aligned}
\end{cases}
\end{equation}
The correction part of the Kalman filter algorithm changes to
\begin{equation}
%\begin{aligned}[b] 
\begin{cases}
{\mathbf{K}_{k}}={{\mathbf{P}}_{k}}^{(p)}{\mathbf{H}_{z}}^{T}(\mathbf{H}_{z}{\mathbf{P}_{k}}^{(p)}\mathbf{H}_{z}^{T} +\mathbf{R})^{-1} \;, \\
{\mathbf{z}_{k}}^{(c)}={\mathbf{z}_{k}}^{(p)}+{\mathbf{K}_{k}}(\mathbf{y}_{k}-\mathbf{H}_{z}\mathbf{z}_{k}^{(p)}) \;,  \\
{\mathbf{P}_{k}^{(c)}}=(\mathbf{I}-\mathbf{K}_{k}\mathbf{H}_{z})\mathbf{P}_{k}^{(p)} \;,
%\end{aligned}
\end{cases}
\end{equation}
where $\mathbf{P}_{k}\in \mathbb{R}^{r\times r}$ and  $\mathbf{K}_{k} \in \mathbb{R}^{r\times P}$ are computed in the reduced-order space. 
The initial predicted values required in starting the iterative algorithm are selected as
\begin{equation}
\begin{cases}
%\begin{aligned}[b]  
{\mathbf{z}_{1}}^{(p)}=\mathbf{\tilde{U}}_{r}^{T}\mathbf{x}_{1} \;,  \\
{\mathbf{P}_{1}}^{(p)}={\mathbf{Q}_{z}} \;.
%\end{aligned}
\end{cases}
\end{equation}   
After applying the Kalman filter, a backward smoother is also applied to produce smoother data~\cite{CrassidisJunkins04}
\begin{equation}
\begin{cases}
%\begin{aligned}[b]  
{\mathbf{K}_{k}^{(s)}}={\mathbf{P}}_{k}^{(c)}{\mathbf{A}_{z}}^{T}({\mathbf{P}_{k+1}}^{(p)})^{-1} \;, \\
{\mathbf{P}_{k}}^{(s)}={\mathbf{P}_{k}}^{(c)}-{\mathbf{K}_{k}}^{(s)}(\mathbf{P}_{k+1}^{(p)}-\mathbf{P}_{k+1}^{(s)})(\mathbf{K}_{k}^{(s)})^{T} \;,\\
{\mathbf{z}_{k}^{(s)}}=\mathbf{z}_{k}^{(c)}-\mathbf{K}_{k}^{(s)}(\mathbf{z}_{k+1}^{(p)}-\mathbf{z}_{k+1}^{(s)}) \;,
%\end{aligned}
\end{cases}
\end{equation}  
where the superscript $s$ denotes the smoothed data. The initial values for the iterative smoothing algorithm are obtained from the last  correction time step at k=N: 
\begin{equation}
\begin{cases}
%\begin{aligned}[b] 
{\mathbf{z}_{N}}^{(s)}=\mathbf{z}_{N}^{(c)} \;,  \\
{\mathbf{P}_{N}}^{(s)}={\mathbf{P}_{N}}^{(c)} \;.
%\end{aligned}
\end{cases}
\end{equation} 
Finally, the reconstructed data in the full-order space is computed as 
 \begin{equation}
\mathbf{V}_{rec}=\mathbf{\tilde{U}}_{r}\mathbf{Z}^{(s)} \;,
\end{equation} 
where $\mathbf{Z}^{(s)}=\begin{bmatrix}\mathbf{z}_{1}^{(s)}  & \mathbf{z}_{2}^{(s)} &\cdots&\mathbf{z}_{N}^{(s)} \end{bmatrix}$. 

\begin{figure}[h!]
\centering
\includegraphics[width=\textwidth,height=\textheight,keepaspectratio]{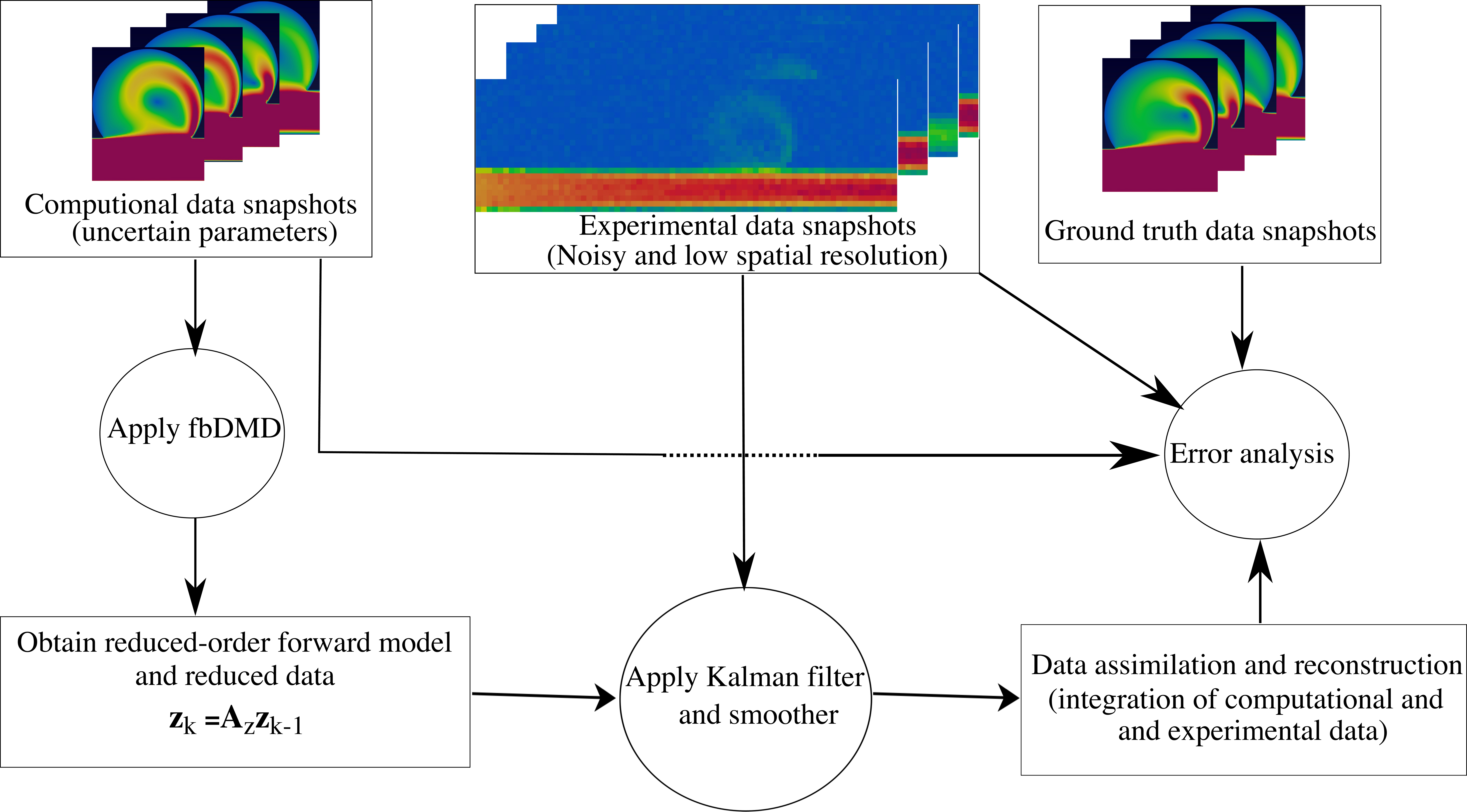}
\caption{An overview of the ROM-KF method for obtaining high-fidelity blood flow data from uncertain computational data and noisy/low-resolution experimental data is sketched.}
\label{fig:methods}
\end{figure}

A summary of the ROM-KF method for obtaining high-fidelity blood flow data is shown in Fig.~\ref{fig:methods}. The reduced-order forward model is obtained by applying fbDMD to the computational data. Then, the reconstructed data are computed by applying ROM-KF on full-order experimental and reduced-order computational datasets. In the end, the accuracy of reconstructed data, computational data, and experimental data are calculated with respect to the ground truth dataset.

\subsection{Noise covariance matrices}
A major challenge in using the linear Kalman filter method is the estimation of the covariance matrices. In this study, we focused on synthetic experimental data; therefore, the noise of the measurement system was given. We propose a simple method to estimate the noise covariance matrix of the computational data without using the ground truth data. The covariance matrices are explained below.

\subsubsection{Covariance matrix for the synthetic experimental data}
The noise of the measurement system is usually given as the characteristic of the system. In this study, the standard deviation ($\sigma_{R}$) of the noise is assumed to be given. Due to the normal distribution assumption of the measurement noise imposed on the synthetic experimental data, the  covariance matrix was assumed to be a diagonal matrix with the variance of the imposed noise on the diagonal elements

\begin{equation}
\mathbf{R} =
\begin{pmatrix}
\sigma_{R}^2 & & \\
& \ddots & \\
& & \sigma_{R}^2 
\end{pmatrix} \;.
\end{equation}

\subsubsection{Covariance matrix for the computational data}

Calculating the covariance matrix of the computational data when there is uncertainty in viscosity and inlet boundary conditions based on common methods such as Monte Carlo simulations is computationally expensive and time-consuming~\cite{Tranetal17}. Here, we proposed a simple method to estimate the standard deviation of the computational data noise/error process. In this method, nine steady simulations with different inlet boundary conditions and viscosities are performed. The average velocity at the inlet for the computational data ($V_{mean}$) and the average viscosity ($\mu_{mean}$) are selected as the same values of the uncertain computational model explained above. Nine steady simulations are computed based on combinations of variations in viscosity $\mu=\{0.71\mu_{mean},\mu_{mean},1.29\mu_{mean}\}$ and inlet boundary condition $V_{inlet}=\{0.85V_{mean},V_{mean},1.15V_{mean}\}$.
The steady simulation result which is computed  with $\begin{bmatrix} V_{inlet}\\ \mu \end{bmatrix}$=$\begin{bmatrix} V_{mean}\\ \mu_{mean} \end{bmatrix}$ is named $\bar{\mathbf{V}}$. Nine simulations were performed by all possible combinations of the above viscosity and inlet flow rates \edit{and with the same outlet boundary conditions as the ground truth models.} Subsequently, the variance of the noise/uncertainty in the computational data is estimated by calculating the variance of the above simulations with respect to the $\bar{\mathbf{V}}$ data. This calculation is done in a point-wise and component-wise fashion where for each point $i$,   $\sigma_{Q_{ix}}$, $\sigma_{Q_{iy}}$, and $\sigma_{Q_{iz}}$ values are obtained. For the purpose of reducing the computational cost and simplicity, the data points are assumed uncorrelated to each other. Therefore, the off-diagonal entries are neglected. Finally, the covariance matrix of the computational data is defined as

\begin{equation}
\mathbf{Q} =
\begin{pmatrix}
\sigma_{Q_{1x}}^2 & &&&& \\
 &\sigma_{Q_{1y}}^2 & &&&\\
 & &\sigma_{Q_{1z}}^2 &&&\\
&&& \ddots && \\
&&&&& \sigma_{Q_{Mx}}^2\\
&&&&&& \sigma_{Q_{My}}^2\\
&&&&&&& \sigma_{Q_{Mz}}^2
\end{pmatrix}
\;,
\end{equation}
 where M is the number of points in the computational data.
\subsection{Error analysis}

To compare the accuracy of the ROM-KF method, computational data, and synthetic experimental data, the normalized root mean square error (NRMSE) at each time-point is computed as
 \begin{equation}
 \textup{NRMSE}(t_{i}) = \frac{1}{\max\limits_{\mathbf{x}_{j}}||V_{ref}(\mathbf{x}_{j},t_{i})||}\sqrt{\frac{1}{K}\sum_{j=1}^{K}\biggl(V(\mathbf{x}_{j},t_{i}) - V_{ref}(\mathbf{x}_{j},t_{i})\biggr)^2}  \;,
\end{equation}
where $K$ is the number of spatial components, $V(\mathbf{x}_{j},t_{i})$ is the modeled or measured velocity at time $t_{i}$ and spatial location $\mathbf{x}_{j}$, and $V_{ref}(\mathbf{x}_{j},t_{i})$ is the ground truth velocity.

\section{Results}
In this section, we evaluate the accuracy of the proposed ROM-KF model in integrating the computational and synthetic experimental data explained above. We also present examples where we do not have any experimental measurement in the near-wall region, but we are interested in reconstructing near-wall blood flow. These examples represent practical scenarios where experimental near-wall blood flow measurement is difficult and often not accurate.

\subsection{Test case 1: Womersley's analytical solution}
\begin{figure}[h!]
\centering
\includegraphics[scale=0.33,keepaspectratio]{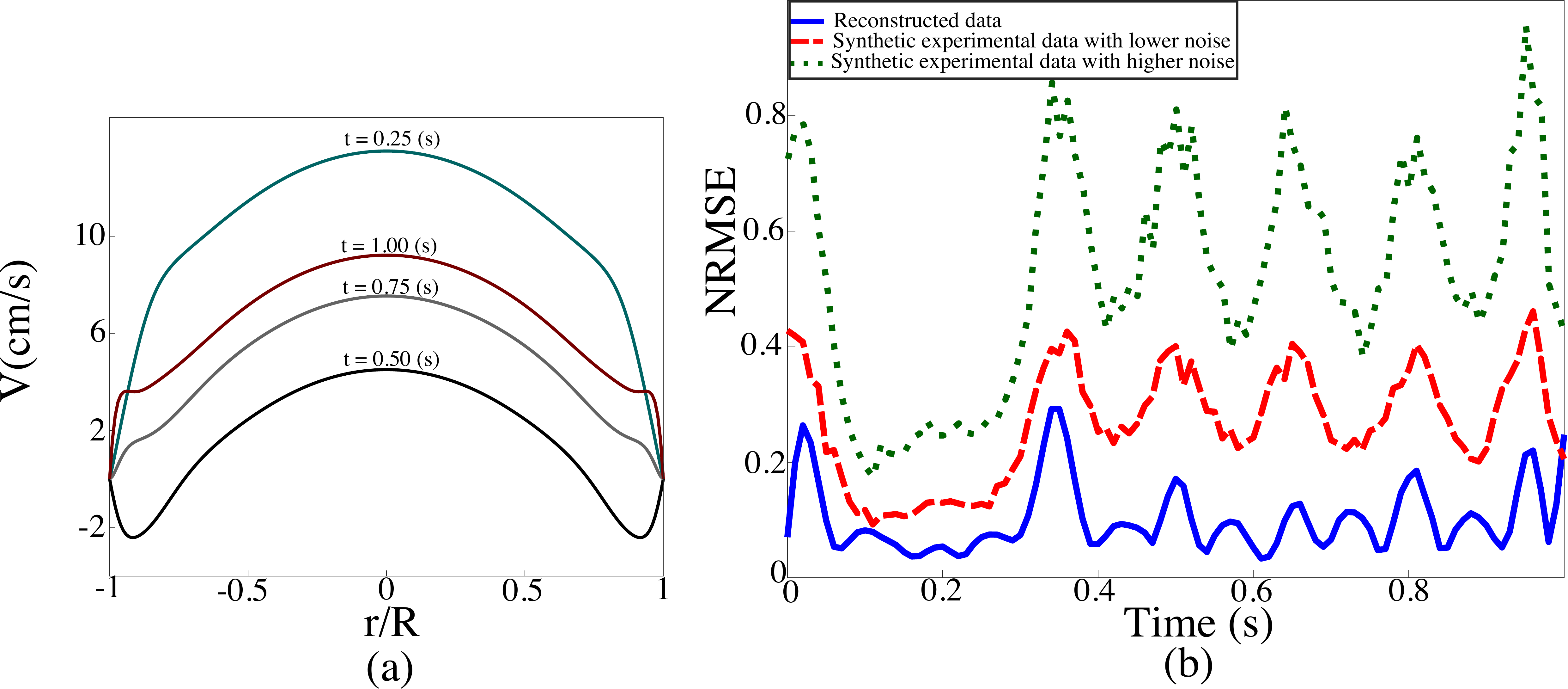}
\caption{(a) The Womersley velocity profiles at several times (0.25, 0.50, 0.75, and 1.00s) are plotted. (b) The NRMSE for reconstructed and synthetic experimental datasets are shown. The solid blue line, dashed red line, and dotted green line represent the NRMSE of reconstructed data, synthetic experimental data with lower noise, and synthetic experimental data with higher noise, respectively. \edit{The maximum time-averaged absolute error of the reconstructed data was 1 cm/s.} }
\label{fig:1DCASE}
\end{figure}

Ground truth data is obtained from Womersley's analytical solution. Synthetic experimental data with two different noise levels (lower noise and higher noise) are constructed as explained in Sec.~\ref{sec:WOMER}. The velocity profiles of the ground truth data at several time-steps are plotted in Fig.~\ref{fig:1DCASE}a. To test our ROM-KF framework, we apply fbDMD to the synthetic experimental data with a lower noise level. The synthetic experimental data with higher noise is considered as measurement data in this test case.  The NRMSE of each dataset with respect to the ground truth data are shown in Fig.~\ref{fig:1DCASE}b. The results show that the reconstructed data has higher accuracy in comparison to the synthetic experimental dataset with lower noise (used in building the fbDMD model) and higher noise (considered as measurements).
 
\subsection{Test case 2: Idealized 2D cerebral aneurysm model}
\begin{figure}[h!]
\centering
\includegraphics[width=\textwidth,height=\textheight,keepaspectratio]{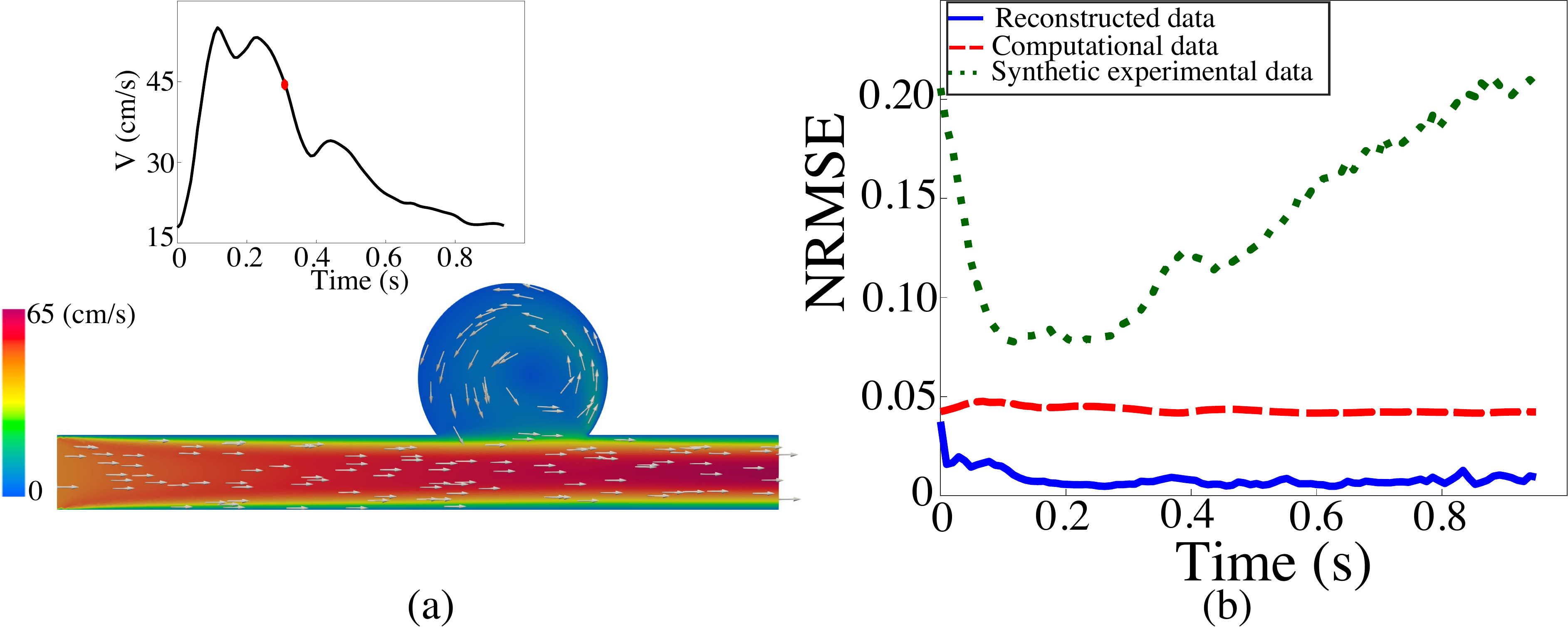}
\caption{(a) The velocity profile of the ground truth data (t=0.3s) \edit{and the pulsatile waveform used as inlet boundary condition are} shown. \edit{The red dot in the waveform shows the time instant of the plotted velocity field.} The visualized velocity vectors are normalized. (b) The solid blue line, the dashed red line, and the dotted green line plot the NRMSE of the reconstructed data, the computational data with uncertain parameters, and the synthetic experimental (mimicking 4D flow MRI) data.  \edit{The maximum time-averaged absolute error of the reconstructed data was 1.8 cm/s.}
}
\label{fig:2DCASE} 
\end{figure}

The ground truth data, the computational data with uncertain parameters, and the synthetic experimental data mimicking 4D-flow MRI  are constructed as explained in Sec.~\ref{sec:2DCer}. The velocity pattern of the ground truth data at t=0.3s is shown in Fig.~\ref{fig:2DCASE}a. The reconstructed data is predicted by applying ROM-KF on the computational data with uncertain parameters and synthetic experimental  data (low resolution and noise). The reconstruction error with respect to the ground truth data is plotted in Fig.~\ref{fig:2DCASE}. As shown, the reconstructed data has consistently lower errors in comparison to the synthetic experimental and uncertain computational datasets throughout all intra-cardiac time-steps.

\subsection{Test case 3:  Patient-specific cerebral aneurysm model}
\begin{figure}[h!]
\centering
\includegraphics[width=\textwidth,height=\textheight,keepaspectratio]{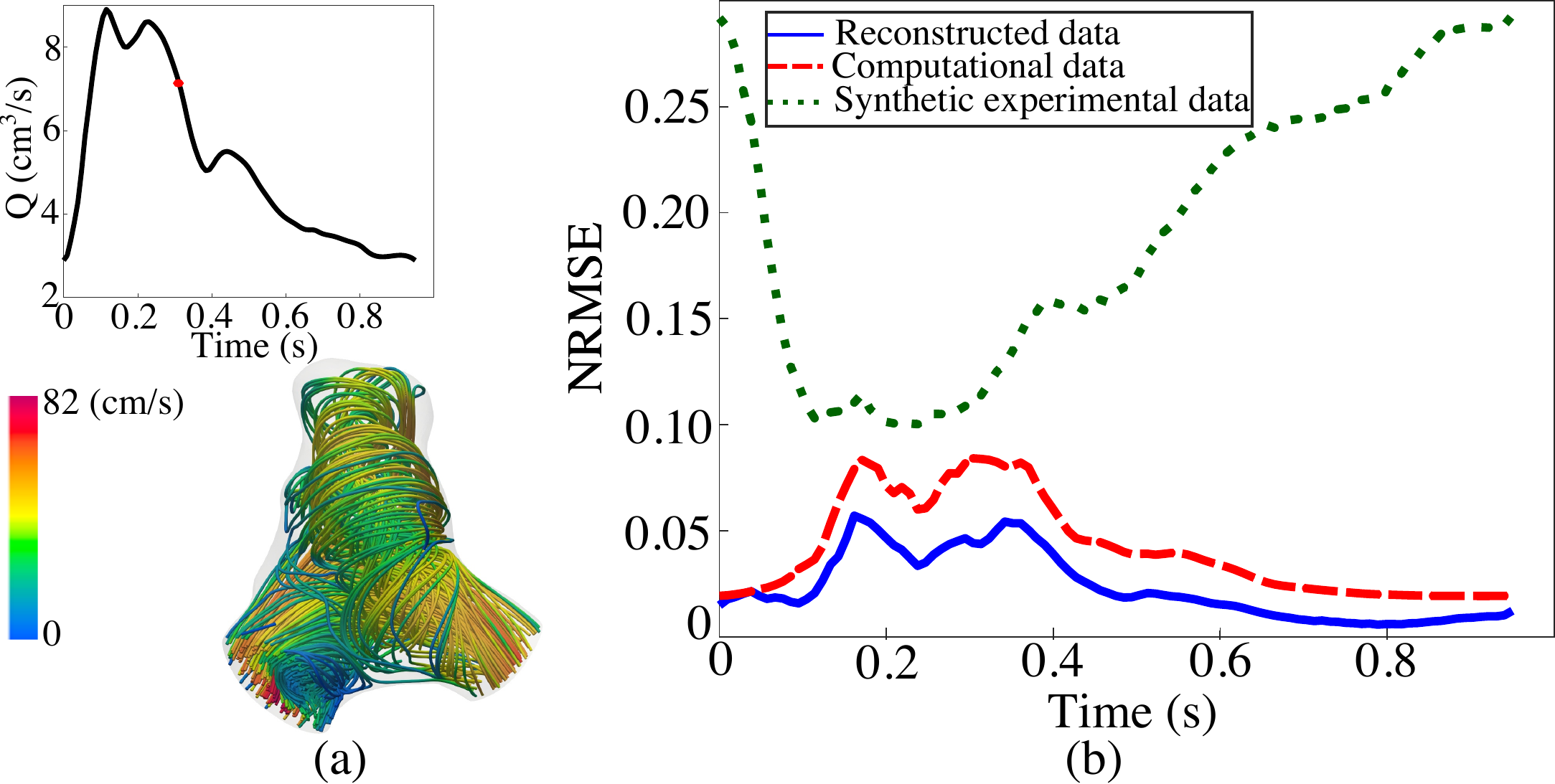}
\caption{(a) The velocity streamlines of the ground truth data (t=0.3s) and \edit{and the pulsatile waveform used as inlet boundary condition are}  shown.  \edit{The red dot in the waveform shows the time instant of the plotted velocity field.} (b) The solid blue line, the dashed red line, and the dotted green line show the NRMSE of the reconstructed data, computational data with uncertain parameters, and synthetic experimental data (noisy and downsampled), respectively.  \edit{The maximum time-averaged absolute error of the reconstructed data was 4.8 cm/s.}
}
\label{fig:3DCASE} 
\end{figure}

In this test case, the computational data with uncertainty in parameters and downsampled noisy experimental data are constructed as explained in Sec.~\ref{sec:3DCer}. The representative velocity streamlines at one snapshot are shown in Fig.~\ref{fig:3DCASE}a. The reconstructed data is obtained by using ROM-KF to integrate the computational and synthetic experimental datasets. The temporal variation of the error is reported in Fig.~\ref{fig:3DCASE}b. The plot shows that during the early time-steps, the ROM-KF reconstructed data has similar accuracy as the uncertain computational data. However, during later stages of the cardiac cycle, the reconstructed data has superior accuracy with respect to the uncertain computational and synthetic experimental datasets.

\subsection{Near-wall flow reconstruction in a 2D idealized cerebral aneurysm} 

\begin{figure}[h!]
\centering
\includegraphics[width=.9\textwidth,height=.9\textheight,keepaspectratio]{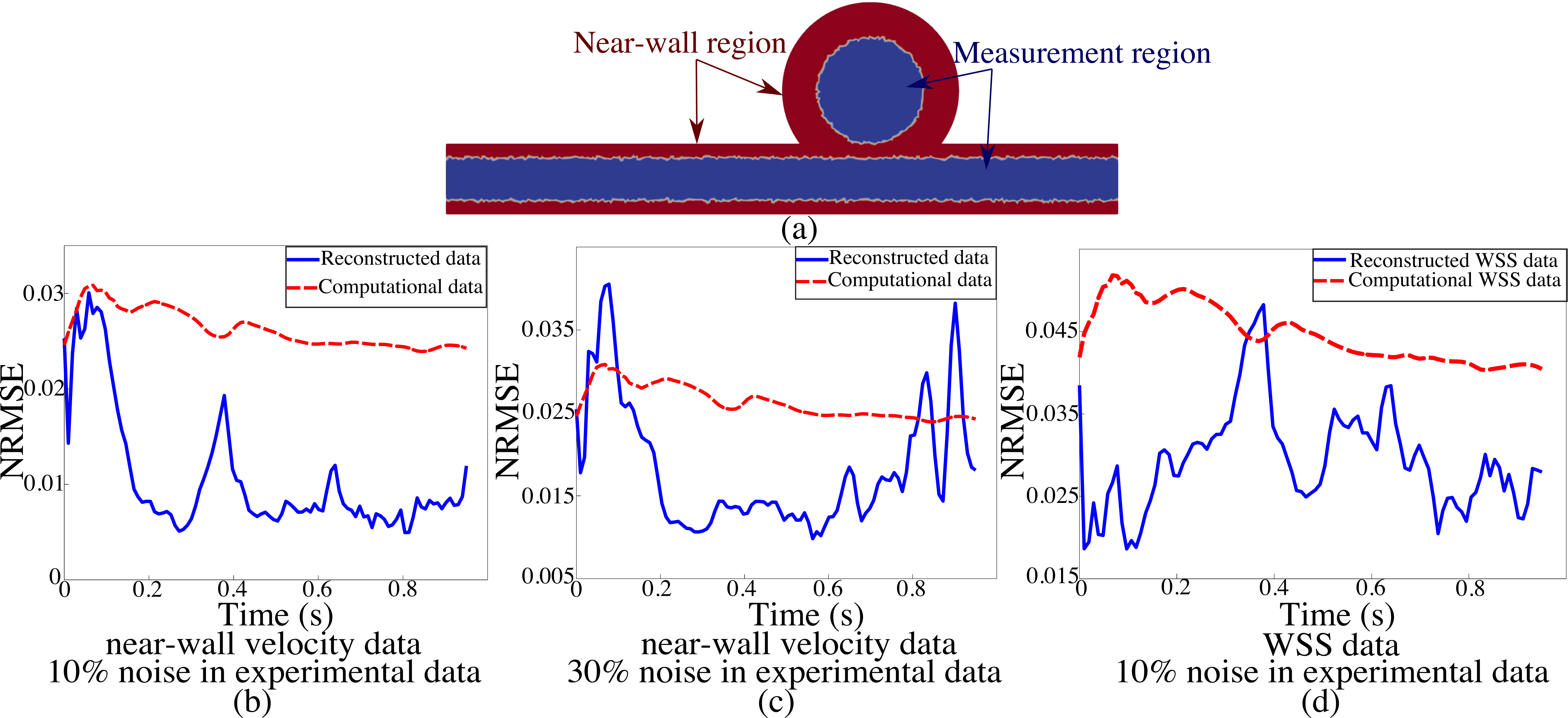}
\caption{(a) The measurement region and the near-wall region where measurement data is lacking are shown for the 2D idealized aneurysm model. The solid blue line and dashed red line represent the NRMSE of the reconstructed \edit{velocity} data and uncertain computational \edit{velocity} data in the near-wall region \edit{(panel b and c) or the reconstructed WSS data and the uncertain computational WSS data (panel d),} respectively. The error \edit{in velocity data} is calculated and plotted for the near-wall region results (hemodynamics data in the red region). The standard deviation of the noise in the experimental data is 10\% of the maximum \edit{global} velocity \edit{of the ground truth data} in (b), 30\% in (c), and \edit{10\% in (d).}}
\label{fig:NEARWALL2D} 
\end{figure}

The same 2D idealized aneurysm model as above is considered. In this example, we discard the near-wall experimental data, mimicking a practical situation where the near-wall flow measurements are not very accurate. The near-wall region (where experimental data is discarded) and the measurement regions are shown in Fig.~\ref{fig:NEARWALL2D}. The ratio of the diameter of the measurement region to the diameter of the parent artery is 0.6. Additionally, the neck of the aneurysm is considered as the near-wall region where measurement data is not included. The ROM-KF method is applied to uncertain computational data and the experimental data in the highlighted measurement region (away from the wall). The NRMSE of the \edit{velocity data} in the near-wall region and \edit{WSS data are} shown in Fig.~\ref{fig:NEARWALL2D}. In these error plots, only the near-wall hemodynamics data are considered (the red near-wall region).  In Fig.~\ref{fig:NEARWALL2D}b, the error is plotted in the near-wall region and compared between the reconstructed data and the computational data.  As shown, by only leveraging the available measurement data away from the wall, the ROM-KF model is capable of reducing the reconstruction error in the near-wall region and produces \edit{near-wall flow results more accurate than the uncertain computational data.} To better understand the effect of measurement data noise \edit{on the results,} Fig.~\ref{fig:NEARWALL2D}c plots the NRMSE in the near-wall region where the noise in the synthetical experimental data is increased to a normally distributed noise with zero mean and 30\% of maximum velocity as standard deviation. We can see that increasing the error in the experimental data reduces the accuracy of the reconstructed data during the early and late stages of the cardiac cycle. \edit{In Fig.~\ref{fig:NEARWALL2D}d, it could be seen that the method improves WSS accuracy in the majority of the time-steps.}

\subsection{Near-wall flow reconstruction in a 3D patient-specific cerebral aneurysm}

\begin{figure}[h!]
\centering
\includegraphics[width=.8\textwidth,height=.8\textheight,keepaspectratio]{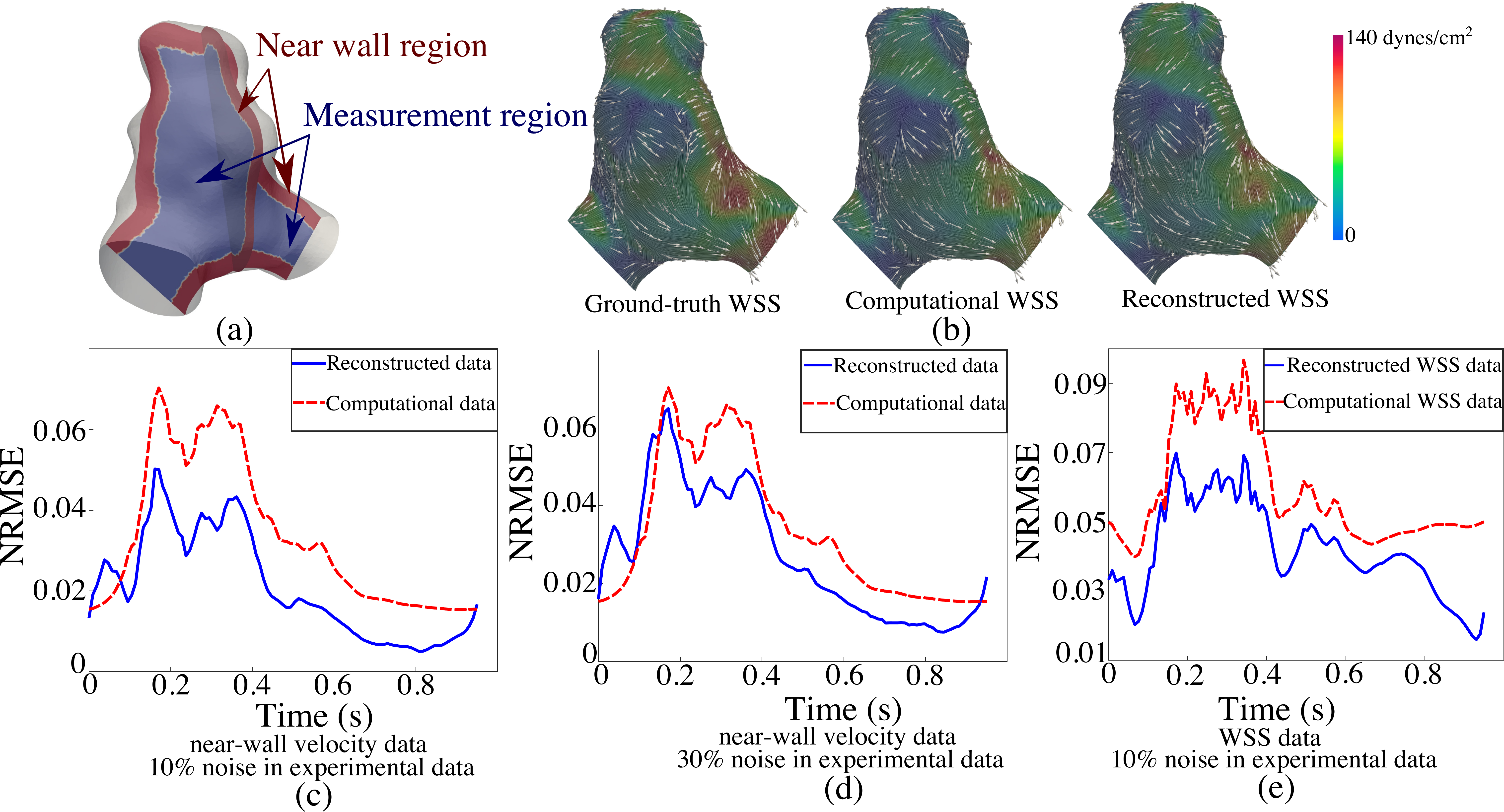}
\caption{(a) The measurement region and the near-wall region where measurement data is lacking are shown for the 3D patient-specific aneurysm model. \edit{(b) WSS streamlines are shown, colored by their magnitude. From left to right: ground truth WSS data, computational WSS data, and reconstructed WSS data at peak systole are shown.} In subsequent panels, the solid blue line and dashed red line represent the NRMSE of reconstructed \edit{velocity/WSS} data and uncertain computational \edit{velocity} data in the near-wall region, respectively. \edit{For the velocity results,} the error is calculated and plotted for the near-wall region (hemodynamics data in the red region). The standard deviation of the noise in the experimental data is 10\% of the maximum \edit{global} velocity \edit{of ground truth data} in (c), 30\% in (d), and \edit{10\% in (e).}}
\label{fig:NEARWALL3D} 
\end{figure}

The 3D cerebral aneurysm model is considered again \edit{with} the near-wall measurement data discarded. The ratio of the diameter of the measurement region to the parent artery diameter is 0.6, as shown in Fig.~\ref{fig:NEARWALL3D}. The reconstructed data is obtained by applying the ROM-KF method on the computational data with uncertain parameters and the given noisy synthetic experimental data, which is only available in the outer-wall measurement region as highlighted in blue in Fig.~\ref{fig:NEARWALL3D}a. \edit{The spatial distribution of the WSS results are shown in Fig.~\ref{fig:NEARWALL3D}b.} The reconstruction error for the near-wall blood flow data is plotted in the figure. In Fig.~\ref{fig:NEARWALL3D}c, 10\% measurement noise is considered. The accuracy of the method in the first time-steps (t$<$0.08s) is less than the uncertain computational data; however, during later time-steps (the majority of the cardiac cycle), the accuracy of the reconstructed data in the near-wall region prevails. In Fig.~\ref{fig:NEARWALL3D}d, the level of noise in the synthetic experimental data is increased to 30\%. In this case, lower accuracy is observed in near-wall flow reconstruction for a longer time (t$<$0.15s), and during later time-steps, the flow reconstruction gains superior accuracy. However, towards the end of the cardiac cycle, the reconstruction error is again inferior to the computational data. \edit{The NRMSE results for WSS are shown in Fig.~\ref{fig:NEARWALL3D}e where improved WSS prediction by the ROM-KD method could be observed.}

%\edit{Also, the NRMSE results of WSS where the noise in the synthetical experimental data with a discarded the near-wall information is a normally distributed noise with zero mean and 10\% of maximum velocity as standard deviation is compared in Fig.~\ref{fig:NEARWALL3D}e. To be consistent with previous assumptions, $\mu=0.04P$ is used to compute ground truth WSS data, $\mu=0.035P$ is used for computation of computational WSS data and reconstructed WSS data. The WSS streamlines are shown in Fig.~\ref{fig:NEARWALL3D}a. at the peak systolic of the inlet waveform for ground truth data, computational data, and reconstructed data. As it is shown in Fig.~\ref{fig:NEARWALL3D}e, by only leveraging the available measurement data away from the wall, the ROM-KF model is capable of producing more accurate WSS results than the uncertain computational data.} 

\begin{figure}[h!]
\centering
\includegraphics[scale=0.6,keepaspectratio]{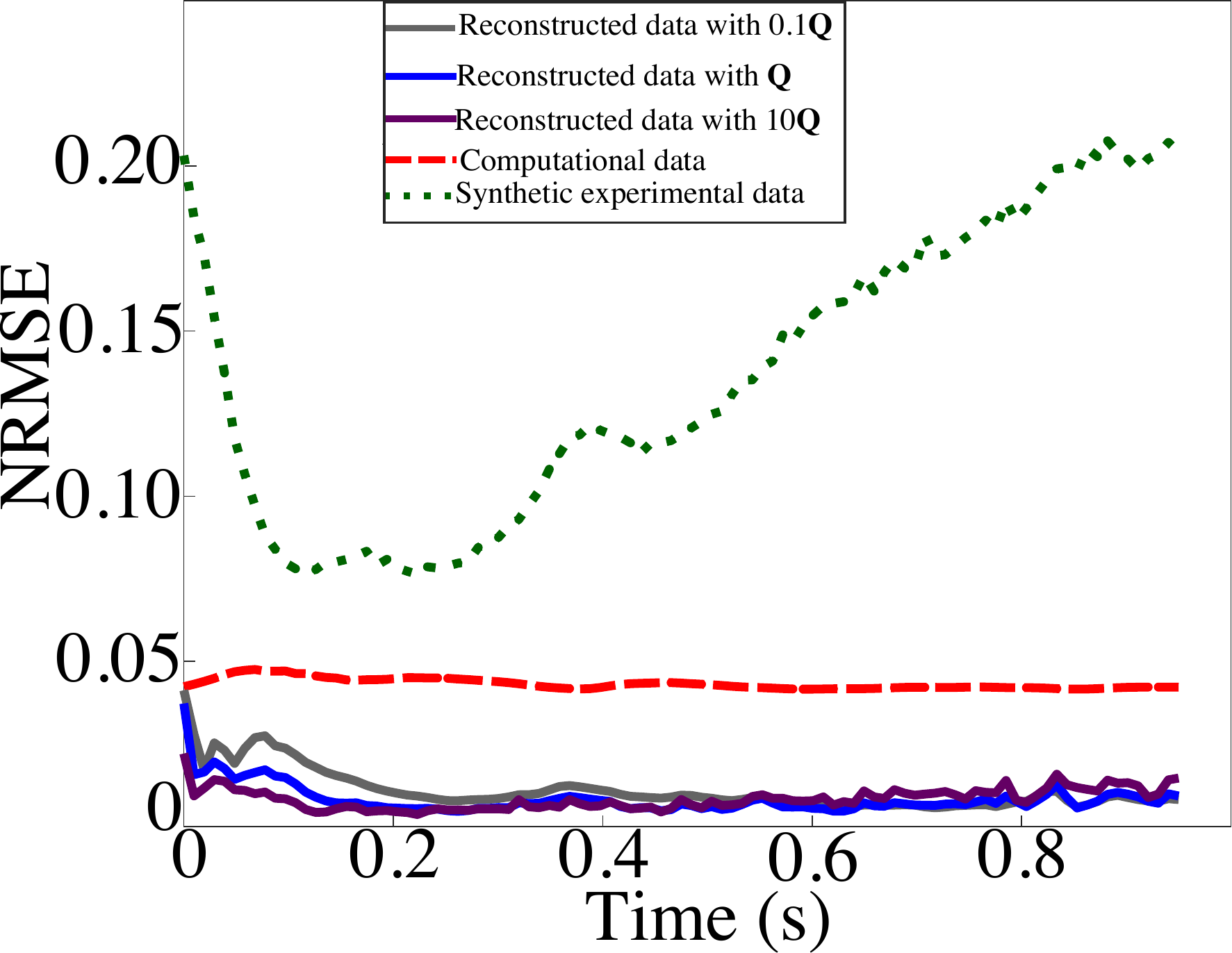}
\caption{The error of three reconstruction with different covariance matrix approximations is compared to the error of the  computational and synthetic experimental data. Three covariance matrices $0.1\mathbf{Q}$, $10\mathbf{Q}$, and $\mathbf{Q}$ are compared where $\mathbf{Q}$ represents the original estimated covariance matrix. }
\label{fig:Sensitivity} 
\end{figure}

\edit{\subsection{Sensitivity analysis of the computational data covariance matrix}\label{sec:SENS}  }
To assess the effect of the computational data covariance matrix approximation on ROM-KF results, the 2D aneurysm model in Sec.~\ref{sec:2DCer} is considered again, and three covariance matrices for the computational data are used. The reconstructed data is predicted by applying ROM-KF on the computational data with uncertain parameters and synthetic experimental data (low resolution and noise) with different covariance matrices for the computational data. Namely, $0.1\mathbf{Q}$ and $10\mathbf{Q}$ values were used in the ROM-KF model and were compared to the results obtained with $\mathbf{Q}$. As shown  in Fig.~\ref{fig:Sensitivity}, the accuracy of three cases over the cardiac cycle is close to each other and remains better than the accuracy of the computational and synthetic experimental data.

\section{Discussion}

We evaluated the power of our proposed ROM-KF method in combining multi-fidelity blood flow data (low-resolution, noisy synthetic experimental data and high-resolution CFD data with uncertainty) \edit{and obtaining} hemodynamic results that are more accurate than either dataset. Our approach reduces the extensive computational cost in performing ensemble Kalman filter methods~\cite{Gaidziketal19} and addresses two outstanding issues in hemodynamics modeling: noise and low spatial resolution in experimental and uncertainty in computational models. We tested our model with idealized 1D/2D and patient-specific 3D blood flow models, where we obtained higher accuracy in our DA results with respect to the multi-fidelity input datasets. Additionally, we demonstrated that our method could increase near-wall hemodynamics accuracy even when experimental data were missing in the near-wall region. 

Our DA framework has important practical implications in high-fidelity patient-specific hemodynamics modeling. If we know the level of noise in our experimental measurement and we can assess the uncertainty of the parameters that are involved in our CFD model, then we can combine our a priori knowledge of noise and uncertainty with multi-modality data to obtain a new dataset that outperforms both the experimental and CFD data in terms of accuracy. In other words, we may think of DA as a two-way benefit for hemodynamics modeling. The measurement data reduces the effect of uncertainties in our computational model parameters, while the computational model allows us to explore spatial and temporal scales that are not accessible by our measurement technique and reduce the inevitable measurement noise. Another promising area is near-wall hemodynamics quantification. We demonstrated that with only measurements away from the wall, our model reduces near-wall hemodynamics error in the computational model. Clinically, there is high interest in correlating WSS with cardiovascular disease; however, high-fidelity quantification of WSS remains a challenge. Interestingly, WSS is one of the most sophisticated hemodynamic parameters due to its multi-faceted role in cardiovascular disease~\cite{Mahmoudietal20} and the challenges in quantifying its rich spatiotemporal vectorial and topological features~\cite{ArzaniShadden16,ArzaniShadden18,Mazzietal21}. As a result, WSS is more sensitive to modeling limitations compared to other hemodynamic measures such as pressure drop. At the same time, current in-vitro and in-vivo experimental techniques (e.g., 4D flow MRI and PIV) have \edit{difficulty} quantifying WSS \edit{compared to} blood flow velocity away from the wall. We have shown that our ROM-KF method leverages the measurements away from the wall to improve near-wall blood flow predictions. 

The success in reconstructing near-wall hemodynamics without near-wall data should not be surprising. Previous work has discussed a close relationship between WSS patterns and flow structures away from the wall~\cite{ArzaniShadden18}. In our ROM-KF algorithm, core region measurement data improve reconstructed data accuracy in the core region during the algorithm's correction steps. Consequently, the forward model improves accuracy in the near-wall region during the algorithm's prediction step using the previous corrected state since the forward operator (DMD model) considers the hemodynamic system's underlying dynamics. Indeed, our previous modal analysis work with DMD has shown a close relationship between WSS and velocity modes in cardiovascular flows~\cite{Habibietal20}. Nevertheless, our algorithm failed to improve near-wall reconstruction error consistently throughout the entire cardiac cycle when the level of noise in the experimental data was sufficiently increased. 

A common observation in our 2D and 3D results was the relatively higher reconstruction error during the start of the cardiac cycle. This is a feature of sequential DA algorithms where during the early time-steps lower accuracy is observed~\cite{Hayase15}. After a sufficient number of time-steps, the algorithm manages to see enough experimental data to be able to correct the effect of uncertainty in the computational model and therefore reduces the reconstruction error. \edit{One approach to solving this problem would be to run ROM-KF for multiple cardiac cycles and consider the reconstructed data from the middle cardiac cycle as high-fidelity hemodynamic data.}  Additionally, DA has the potential to stabilize reduced-order models such as DMD when unstable modes exist and therefore improving DMD prediction. 

%

%An interesting and perhaps a hidden feature of the ROM-KF model is the effect of data assimilation on 

One of the challenges in using our ROM-KF method was approximating the covariance matrices for the computational model. In our approach, the covariance matrix depends on the propagation of the uncertainty in viscosity and inlet boundary conditions.  To approximate the covariance matrix, we have proposed a simple method based on a very limited number of steady simulations. This  simple approach was selected to reduce the high computational cost involved in rigorous uncertainty quantification with Monte Carlo type methods~\cite{Tranetal17,Seoetal20}, which are also used in the ensemble Kalman filter approach. However, it should be noted that even with more rigorous uncertainty quantification, the Kalman filter framework is assuming that the error distribution is Gaussian, which is likely not the case in practice. Therefore, there will always be some discrepancy between the true and assumed statistics, even with perfect information. To assess the influence of the simple covariance matrix calculation method on the \edit{predictions,} the results of three covariance matrices were compared in  Fig.~\ref{fig:Sensitivity}, where it was observed that uniformly scaling the covariance matrix by an order of magnitude does not seem to notably affect the predictions.

A major limitation of our study is that we have used synthetic experimental data. The use of synthetic experimental data provides an ideal setting to validate and quantify the accuracy of our model. Ideally, one would select low-resolution in-vivo data as the experimental data, perform DA with CFD, and assess the accuracy of the results with in-vitro measurements such as PIV, which is considered the gold-standard method in experimental fluid mechanics~\cite{Raffeletal18} and is commonly used in hemodynamics research~\cite{Charonkoetal09,Keshavarzetal14,Mederoetal18,HatoumDasi18}. However, unfortunately, this approach does not perform a real validation. In our DA method, we are performing ultra-high-resolution CFD simulation (quadratic high order elements), and therefore the numerical error is sufficiently small.  The major error in such CFD simulations comes from uncertainty in patient-specific parameters. In-vitro PIV has the exact same issue as CFD. The information that we use in setting up CFD and PIV experiments are the same (boundary conditions and constitutive properties), questioning the utility of experimental validation in this case. Therefore, we generated synthetic experimental data where we have an exact ground truth simulation \edit{to} validate our model. Of course, the success of our method with real-world data remains to be investigated and is expected to be challenging. The difficulty here is due to the lack of a clear validation benchmark. Also, assessing the covariance matrix in real-world experimental data is not trivial. We should point out that under the scenario where we exactly know the CFD simulation parameters, but we have a low-resolution and numerically dissipative CFD solver where numerical discretization error is prominent, we can assess our DA method by leveraging PIV data as benchmark for validation. \edit{Our work has additional limitations. We did not consider non-Newtonian rheology models, but our ROM-KF model is data-driven and could be readily applied to CFD data obtained from more complex models. The use of ROM within our model will generate error in reconstructing the fine-scale flow structures depending on the complexity of the flow.}

Another limitation of most DA methods is that they are not guaranteed to satisfy the governing equations and conservation laws. A possible remedy to this is to use physics-informed neural networks (PINN)~\cite{Raissietal19}. In recent work, PINN has been used for superresolution and denoising of 4D flow MRI~\cite{Fathietal20b} and synthetic low-quality hemodynamics data~\cite{GaoSunWang20}. We may also think of this approach as a deep learning DA strategy where we use PINN to find spatiotemporal hemodynamics. In such PINN problems, the low-resolution experimental data is the data loss in our neural network and the mathematical equations (physics loss) have replaced our computational or reduced-order DMD model. In general, machine learning and DA are closely related, and they are different approaches to inverse modeling and optimization~\cite{Geer21}. For instance, variational DA and neural networks have mathematical similarities~\cite{Abarbaneletal18}. Traditionally, DA could integrate physical laws more readily in its framework (the prediction step); however, with the advent of PINNs, it is now possible to easily incorporate physical laws in machine learning~\cite{Raissietal19}. 

There are other methods beyond Kalman filter and neural networks for fusing experimental and numerical data. One approach is to combine reduced-order modeling with computational and experimental data in a machine learning framework where a classical optimization problem (e.g., compressed sensing) is used to merge multi-modality data. This approach has been used in merging hemodynamics data~\cite{Bakhshinejadetal17,Fathietal18,ArzaniDawson20}. We may think of this as a machine learning reduced-order model (ML-ROM) framework~\cite{Bruntonkutz19,ArzaniDawson20}. Full-order optimization has also been \edit{utilized} in merging multi-modality blood flow data~\cite{Klemensetal20,Togeretal20}. It is also possible to leverage neural networks for the optimization problem involved in combining ROM models from different datasets~\cite{Casasetal20}. Modeling the effect of discrepancy in experimental and computational data by a feedback source term in the Navier-Stokes equations is a simple method that has been used in merging CFD and 4D flow MRI data~\cite{Annioetal19}. Other methods such as Gaussian process regression~\cite{Perdikarisetal17} and Monte Carlo sampling~\cite{Peherstorferetal18} could be used in multi-fidelity modeling and have been successfully applied to 1D blood flow modeling in the pulmonary circulation~\cite{Paunetal20}. The relative success of all of these methods in the context of 3D hemodynamics modeling remains to be investigated. 

In summary, recent advances in data-driven modeling and scientific machine learning have a high potential to transform patient-specific cardiovascular flow modeling~\cite{ArzaniDawson20}. Such methods not only improve our fundamental understanding of blood flow physics in complex diseased arteries but can also change the way we model hemodynamics by paving the way for appropriate blending of multi-modality data with the ultimate goal of high-fidelity patient-specific cardiovascular flow modeling.

 \section{Conclusion}            
In this paper, a new method called reduced-order model Kalman filter (ROM-KF) was proposed for obtaining high fidelity blood flow data by merging multi-fidelity hemodynamics data. The proposed ROM-KF method is a non-intrusive, parameter-free, and computationally cheap method. The proposed method was tested with different test cases demonstrating its success. Additionally, it was demonstrated that the method could overcome a common difficulty in experimental modeling of cardiovascular flows where near-wall hemodynamics quantification is difficult.

\section{Appendix}
\subsection{DMD}

DMD is an equation-free technique that can provide a linear discrete equation relating different snapshots of data (${\mathbf{x}_{k+1}}=\mathbf{A}\mathbf{x}_{k}$). In DMD, first, the snapshots of data are arranged into $\mathbf{X}$ and $\mathbf{X'}$ as $\mathbf{X'}=\begin{bmatrix}\mathbf{x}_{2} & \mathbf{x}_{3}&\cdots&\mathbf{x}_{N}\end{bmatrix}$ and $\mathbf{X}=\begin{bmatrix}\mathbf{x}_1 & \mathbf{x}_2 & \cdots & \mathbf{x}_{N-1}\end{bmatrix}$. To find a linear relation between $\mathbf{X}$ and $\mathbf{X'}$ ($\mathbf{X'} = \mathbf{A}\mathbf{X}$) the operator $\mathbf{A}$ can be computed as: 

{\begin{equation}
\mathbf{A} = \mathbf{X'}\mathbf{X}^{+}\;,
\label{equ:DMD}
\end{equation}   
where $\mathbf{X}^{+}$ is the pseudo-inverse of $\mathbf{X}$.  The best fit linear model $\mathbf{A}$  is obtained by minimizing the Frobenius norm of data reconstruction
$||\mathbf{X'}-\mathbf{A}\mathbf{X}||_{F}$. One objective of DMD is often to find the eigenvectors and eigenvalues of the underlying system $\mathbf{A}$. Also, DMD could find a compact (reduced-order) matrix $\mathbf{\tilde{A}}$  which can approximate the full matrix $\mathbf{{A}}$. The DMD algorithm proceeds as follows~\cite{Tuetal13}:

\begin{enumerate}[leftmargin=*,labelsep=4.9mm]

  \item Given data, construct snapshot matrices $\mathbf{X}$ and $\mathbf{X'}$.\
   \item Compute the singular value decomposition (SVD) of  $\mathbf{X}$ and obtain the truncated decomposition $\mathbf{X}\approx \mathbf{{U}}\mathbf{{\Sigma}{V}^{*}} $ where we only keep the first  $r$ singular values. The  truncation value $r$ is obtained based on hard thresholding~\cite{GavishDonoho14}.
   \item The matrix  $\mathbf{A}$ can be obtained by using the pseudo-inverse of $\mathbf{X}$:
   \begin{equation}
\mathbf{A} = \mathbf{X'}\mathbf{V}\mathbf{{\Sigma}^{-1}\mathbf{U}^{*}}\;.
\label{equ:DMD}
\end{equation}   
It is more efficient to compute the reduced-oder version of matrix $\mathbf{A}$
 
 \begin{equation}
\mathbf{\tilde{A}} =\mathbf{U}^{*}\mathbf{{A}}\mathbf{U}=\mathbf{U}^{*}\mathbf{X'}\mathbf{V}\mathbf{{\Sigma}^{-1}}\;.
\label{equ:DMD}
\end{equation}  
 \item Compute the spectral decomposition of $\mathbf{\tilde{A}}$ as $\mathbf{\tilde{A}W=W\Lambda}$ to find its eigenvectors ($\mathbf{W}$) and eigenvalues ($\mathbf{\Lambda}$). 
 \item Extract the dynamic modes of the operator $\mathbf{A}$ 
\begin{equation}
 \mathbf{\Phi}=\mathbf{X'V\Sigma^{-1}}\mathbf{W} \;.
\end{equation}

\end{enumerate}

\subsection{fbDMD}
%The fbDMD approach is based on a similar idea to the Kalman smoother method that we used where a backward pass to the data is performed for smoothing purposes
To overcome noise effects in obtaining the linear model, the forward-backward DMD (fbDMD) is applied to the computational dataset. The following algorithm is used to obtain the linear discrete model and the reduced-order subspace of the snapshot matrix~\cite{Dawsonetal16}:

\begin{enumerate}[leftmargin=*,labelsep=4.9mm]
  \item Given snapshots of data, construct the snapshot matrices $\mathbf{X'}=\begin{bmatrix}\mathbf{x}_{2} & \mathbf{x}_{3}&\cdots&\mathbf{x}_{N}\end{bmatrix}$ and $\mathbf{X}=\begin{bmatrix}\mathbf{x}_1 & \mathbf{x}_2 & \cdots & \mathbf{x}_{N-1}\end{bmatrix}$.

  \item Compute the SVD of the matrix $\mathbf{X}$ to get $\mathbf{X}=\mathbf{\tilde{U}}\mathbf{\tilde{\Sigma}} \mathbf{\tilde{V}^*}$. Subsequently, based on the SVD results and the truncation value $r$, the truncated SVD is computed as $\mathbf{X}\approx \mathbf{\tilde{U}}_{r}\mathbf{\tilde{\Sigma}}_{r} \mathbf{\tilde{V}}_{r}^*\label{equ:POD} $. The truncation value $r$ is selected based on  the Singular Value Hard Thresholding method~(SVHT) applied on $\mathbf{X}$~\cite{GavishDonoho14}.

  \item Compute a reduced-order approximation of the forward linear operator $\mathbf{\tilde{A}}_{fwd}=\mathbf{\tilde{U}}_{r}^{*}\mathbf{X^{'}}\mathbf{\tilde{V}}_{r}\mathbf{\tilde{\Sigma}}_{r}^{-1}$.
  
    \item Compute the SVD of the matrix $\mathbf{X^{'}}$ to get $\mathbf{X^{'}}=\mathbf{\tilde{U}}_{bwd}\mathbf{\tilde{\Sigma}}_{bwd} \mathbf{\tilde{V}}_{bwd}^*$. Subsequently, based on the SVD results and the truncation value ${r}$, the truncated SVD will be $\mathbf{X^{'}}\approx \mathbf{\tilde{U}}_{r,bwd}\mathbf{\tilde{\Sigma}}_{r,bwd} \mathbf{\tilde{V}}_{r,bwd}^*$. Alternatively, to reduce the computational cost of the fbDMD algorithm,  SVD of all data ($\mathbf{Y}=\begin{bmatrix}\mathbf{x}_1 & \mathbf{x}_2 & \cdots & \mathbf{x}_{N}\end{bmatrix}$) could be used rather than two separate decompositions to compute the reduced-order forward/backward linear operator.
      \item Compute a reduced-order approximation of the backward linear operator. 
     \begin{equation} 
        \mathbf{\tilde{B}}_{bwd}=\mathbf{\tilde{U}}_{r,bwd}^{*}\mathbf{X}\mathbf{\tilde{V}}_{r,bwd}\mathbf{\tilde{\Sigma}}_{r,bwd}^{-1} \;.
      \end{equation}

\item Compute a reduced-order approximation of the linear operator $\mathbf{\tilde{A}}$
\begin{equation}
\mathbf{\tilde{A}}=(\mathbf{\tilde{A}}_{fwd}\mathbf{\tilde{B}}_{bwd}^{-1})^{1/2} \;.
\end{equation}  

\item Compute the eigenvalues $\mathbf{\Lambda}$ and eigenvectors $\mathbf{W}$ of $\mathbf{\tilde{A}}$ based on the spectral decomposition of $\mathbf{\tilde{A}}$ as $\mathbf{\tilde{A}W=W\Lambda}$.
\item Extract the dynamic modes of the operator $\mathbf{A}$
\begin{equation}
 \mathbf{\Phi}=\mathbf{X'}\mathbf{\tilde{V}}_{r}\mathbf{\tilde{\Sigma}}_{r}^{-1}\mathbf{W} \;.
\end{equation} 

\end{enumerate}

 \bibliographystyle{unsrt}
 \bibliography{DA}

\end{document}